\documentclass[fleqn,usenatbib]{mnras}

\usepackage{newtxtext,newtxmath}

\usepackage[T1]{fontenc}

\DeclareRobustCommand{\VAN}[3]{#2}
\let\VANthebibliography\thebibliography
\def\thebibliography{\DeclareRobustCommand{\VAN}[3]{##3}\VANthebibliography}
\usepackage{xcolor}

\usepackage{graphicx}	
\usepackage{amsmath}	

\title[Interpretation of the data on magnetars]{An alternative interpretation of magnetars' traits deduced from the observational data on their outburst fluxes and spectra}

\author[H. Ardavan]{
Houshang Ardavan\thanks{Email address: ardavan@ast.cam.ac.uk}\\
Institute of Astronomy, University of Cambridge,
Madingley Road, Cambridge CB3 0HA, UK}

\date{Accepted XXX. Received YYY; in original form ZZZ}

\pubyear{2022}

\begin{document}
\label{firstpage}
\pagerange{\pageref{firstpage}--\pageref{lastpage}}
\maketitle

\begin{abstract}
By applying the Efron-Petrosian method to the fluxes $S$ and distances $D$ of the magnetars listed in the Magnetar Outburst Online Catalogue, we show that the observational data are consistent with the dependence $S\propto D^{-3/2}$, which characterizes the emission from the superluminally moving current sheet in the magnetosphere of a non-aligned neutron star, at substantially higher levels of significance than they are with the dependence $S\propto D^{-2}$. This result agrees with that previously obtained by an analysis of the data in the McGill Online Magnetar Catalog and confirms that, contrary to the currently prevalent view, magnetars' X-ray luminosities do not exceed their spin-down luminosities.  The X-ray spectra of magnetars, moreover, are congruous with the spectral energy distribution (SED) of a broadband non-thermal emission mechanism identical to that at play in rotation-powered pulsars: we show that the SED of the caustics that are generated in certain privileged directions by the magnetospheric current sheet single-handedly fits the observed spectra of 4U 0142+61, 1E 1841-045 and XTE J1810-197 over their entire breadths.  Magnetars' outbursts and their associated radio bursts are predicted to occur when, as a result of large-scale timing anomalies (such as glitches, quakes or precession), one of the privileged directions along which the radiation from the current sheet decays more slowly than predicted by the inverse-square law either swings past or oscillates across the line of sight.
\end{abstract}


\section{Introduction}
\label{sec:introduction}

The radiation field that arises from the superluminal motion of the distribution pattern of the current sheet in the magnetosphere of a non-aligned neutron star was analysed in~\citet{Ardavan2021} and was shown to account for a number of salient features of the pulsar emission: its brightness temperature, polarization, spectral distribution and profile with microstructure and with a phase lag between the radio and gamma-ray pulses.  The predicted characteristics of this radiation were subsequently confronted with observational data in~\citet{Ardavan2023a,Ardavan2023b,Ardavan2024b,Ardavan2024a}: with the Fermi-LAT data on the fluxes of gamma-ray pulsars~\citep{Abdollahi2022}, with the data on the curved features of the spectra of radio pulsars catalogued by~\citet{Swainston} and with the data on the gamma-ray spectra of the Crab, Vela and Geminga pulsars~\citep{Abdo2010,Aleksic2011,Ansoldi2016,Abdo2013,MAGIC2020,HESS2023}.  (For a heuristic account of the theory underlying the broadband radiation that is generated by the magnetospheric current sheet of a non-aligned neutron star, see~\citealt{Ardavan2022b}.)

The present paper is concerned with the fluxes and spectra of the emission from the particular set of X-ray emitting neutron stars, classed as magnetars, that are deemed too luminous to be rotation-powered: a notion that is based on the long spin periods and the relatively large spin-down rates of this set of neutron stars and the assumption that the decay of their observed emission with distance would necessarily obey the inverse-square law~\citep{Mereghetti2015,Turolla2015,Kaspi2017,Borghese2019,Esposito2021}.  This notion has led, in turn, to the widely held view that such neutron stars are endowed with an ultra-strong magnetic field whose dissipation and instabilities power the emission received from them~\citep{Duncan1992} and underlie the flaring activity that characterizes their emission~\citep{Beloborodov2016}.  To see whether these views are supported by observations, we analyse here the data on the outburst fluxes and distances of the magnetars listed in the Magnetar Outburst Online Catalogue~\citep{MOOC2018}.  We also examine the observed spectral energy distribution (SED) of the X-ray emission from a selection of representative magnetars to determine whether these distributions, which are normally fitted with a combination of several power-law and black-body spectra~\citep[e.g.][]{Rea2008,An2013,BorgheseXTE}, are congruous with the SED of the broadband non-thermal emission mechanism by which the magnetospheric current sheet radiates. 

Our analysis of the distance dependence of the outburst fluxes of magnetars (Section~\ref{sec:data}) is based on the requirement that the observed luminosities and distances of any given population of an astronomical object should represent two {\it independent} sets of random variables~\citep[e.g.][]{Ivezic2020}.  To determine the exponent $\alpha$ in the dependence $S\propto D^{-\alpha}$ of the flux densities $S$ of magnetars on their distances $D$ for which this requirement is satisfied, we here apply a generalization of the classical tests of independence~\citep{Hajek} known as the Efron-Petrosian method: a method that takes the incompleteness of the tested data sets into account~\citep{EF1992, Maloney1999,Petrosian2002,Bryant, Desai2023}.  We find that the observational data are consistent with $\alpha=-3/2$, which characterizes the emission from the superluminally moving current sheet in the magnetosphere of a non-aligned neutron star~\citep{Ardavan2021,Ardavan2022b,Ardavan2023a}, at substantially higher levels of significance than they are with $\alpha=-2$ (Section~\ref{subsec:results}).  The ratio of X-ray to spin-down luminosity that is estimated by using $\alpha=-3/2$ instead of $\alpha=-2$ turns out to be appreciably less than $1$ for all known magnetars (Section~\ref{sec:Xrayluminosity}).

Having thus demonstrated that the observational data do not uphold the notion that magnetars' X-ray luminosities exceed their spin-down luminosities (Section~\ref{sec:test}), we next examine the data on the X-ray spectra of the following extensively studied magnetars: 4U0142+61, 1E1841-045 and XTE J1810-197~\citep{denHartog2008, FermiMagnetars,MagicMagnetars,Hascoet2014,An2013,BorgheseXTE}.  We begin (in Section~\ref{subsec:spectrum}) by recapitulating the analytic expression for the SED of the most tightly focused component of the radiation that is emitted by the magnetospheric current sheet~\citep{Ardavan2024b}.  We then specify the values of the free parameters in the described SED for which this expression best fits the data on the X-ray spectra of the magnetars 4U0142+61, 1E1841-045 and XTE J1810-197 (Section~\ref{subsec:fits}).  The specified values of the fit parameters will be used, in conjunction with the results of the theoretical analysis in~\citet{Ardavan2021}, to determine certain attributes of the central neutron stars of these magnetars and their magnetospheres in Section~\ref{subsec:connection}.  We will see that all observed features of the spectra of these three magnetars can be described by the SED of a single non-thermal emission mechanism: an emission mechanism identical to that responsible for the spectrum and polarization of rotation-powered pulsars~(\citealt{Ardavan2021}, section~5.1, and~\citealt{Ardavan2024a}) to which the observed polarization of magnetars' X-ray signals~\citep{Zane2023,Heyl2024,Taverna2024} is intrinsic (Section~\ref{subsec:connection}).

On the basis of the results reported in Sections~\ref{sec:test} and~\ref{sec:spectra}, and the fact that the flaring activity of magnetars is often accompanied by timing anomalies such as glitches, mode changes or quakes~\citep{Archibald2020,Champion2020,Lower2023,Younes2023,Tsuzuki2024,Hu2024,Fisher2024}, an alternative interpretation of magnetars' outbursts and their association with fast radio bursts is put forward in Section~\ref{sec:conclusion}. 

\section{Energetic requirements of the X-ray emission from magnetars}
\label{sec:test}

\subsection{The observational data on outburst fluxes of magnetars}
\label{sec:data}

\begin{figure*}
\centerline{\includegraphics[width=18cm]{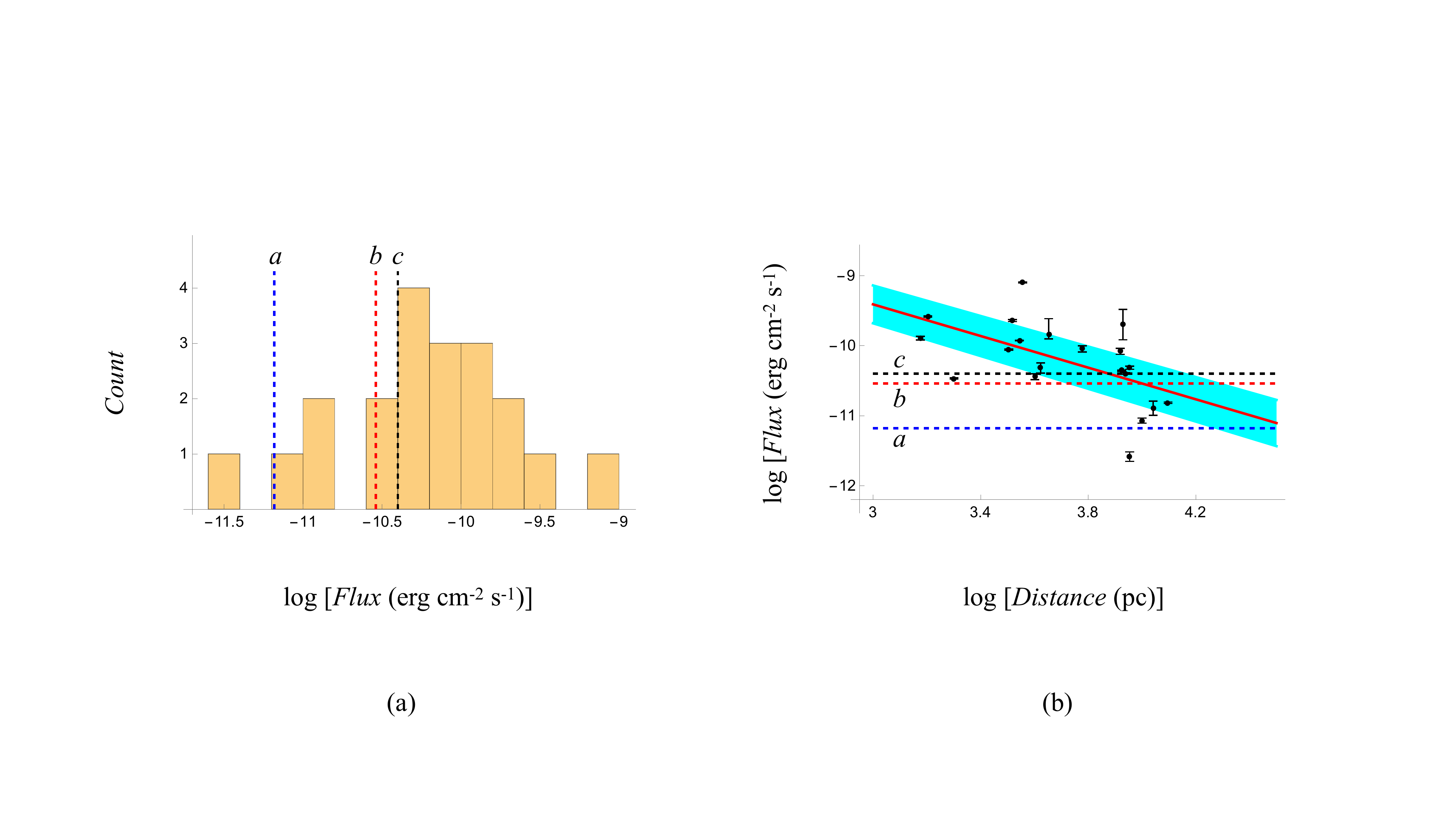}}
\caption{(a) Histogram of the $20$ magnetars whose outburst fluxes are listed by~\citet{MOOC2018}.  The broken lines $a$, $b$ and $c$ designate the flux thresholds $\log S_{\rm th}=-11.18$, $-10.54$ and $-10.4$, in units of erg cm${}^{-2}$ s${}^{-1}$, respectively.  (b) Distribution of logarithm of flux versus logarithm of distance for these magnetars.  The red solid line, whose slope has the value $-1.13\pm0.05$, and the area coloured cyan show the least-squares fit of a straight line to the plotted data points and its uncertainty band, respectively.  The broken lines $a$, $b$ and $c$ designate the same flux thresholds as those shown in part (a). } 
\label{UM1}
\end{figure*}

Tables~3 and 5 of~\citet{MOOC2018} list the distances and the unabsorbed X-ray fluxes of 20 magnetars which show major outbursts or variations in their persistent emission.  If we replace the fluxes corresponding to those $8$ of these $20$ magnetars for which two separate outbursts are listed by their average values, we obtain the histogram shown in Fig.~\ref{UM1}a.  Logarithm of the X-ray flux (in units of erg cm$^{-2}$ s$^{-1}$) of each magnetar is plotted versus the logarithm of its distance (in units of pc) in Fig.~\ref{UM1}b, in which we have averaged the errors listed for the magnetars with two separate outbursts and have set the errors in the cases of SGR 1745-2900 and PSR J1846-0258 (for which no uncertainties are listed in~\citealt{MOOC2018}) equal to a tenth of their fluxes.  

The least-squares fit of a straight line to the plotted data points and its uncertainty band are respectively depicted by the red solid line (whose slope has the value $-1.13\pm0.05$) and by the area coloured cyan in Fig.~\ref{UM1}b.  The dashed lines $a$, $b$ and $c$ in Figs~\ref{UM1}a and~\ref{UM1}b, which correspond to the values $-11.18$, $-10.54$ and $-10.4$ of the logarithm of the flux in units of erg cm${}^{-2}$ s${}^{-1}$, respectively, each designate a flux threshold (i.e. a flux below which the plotted data set may be regarded as incomplete).  The number of elements of the shown data set that are excluded by the thresholds $a$, $b$ and $c$ are $1$, $4$ and $6$, respectively.  

The significant departure of the value of the slope of the line fitted to the data in Fig.~\ref{UM1}b from that predicted by the inverse-square law, i.e.\ from $-2$, indicates that these data cannot be regarded as complete.  Whether this departure can be attributed entirely to the incompleteness of the present data set can be ascertained by means of the Efron-Petrosian method~\citep{EF1992}.  

To evaluate the Efron-Petrosian statistic $\tau$ for a given decay exponent $\alpha$ and a given flux threshold $S_{\rm th}$, we here follow the same procedure as that detailed in~\citet[][section 3.1]{Ardavan2023a}.  The hypothesis of independence of luminosities and distances of magnetars is rejected if the resulting value of $\tau$ renders the quantity
\begin{equation}
p={\rm erfc}\left(\frac{\vert\tau\vert}{\sqrt{2}}\right)
\label{E7}
\end{equation}
smaller than an adopted significance level between $0$ and $1$, where erfc denotes the complementary error function (see equation 7 of~\citealt{Ardavan2023a}). When $\tau$ equals $0$, for instance, $p$ assumes the value $1$ and so the hypothesis of independence of luminosity and distance cannot be rejected at any significance level.

\begin{figure}
\centerline{\includegraphics[width=16cm]{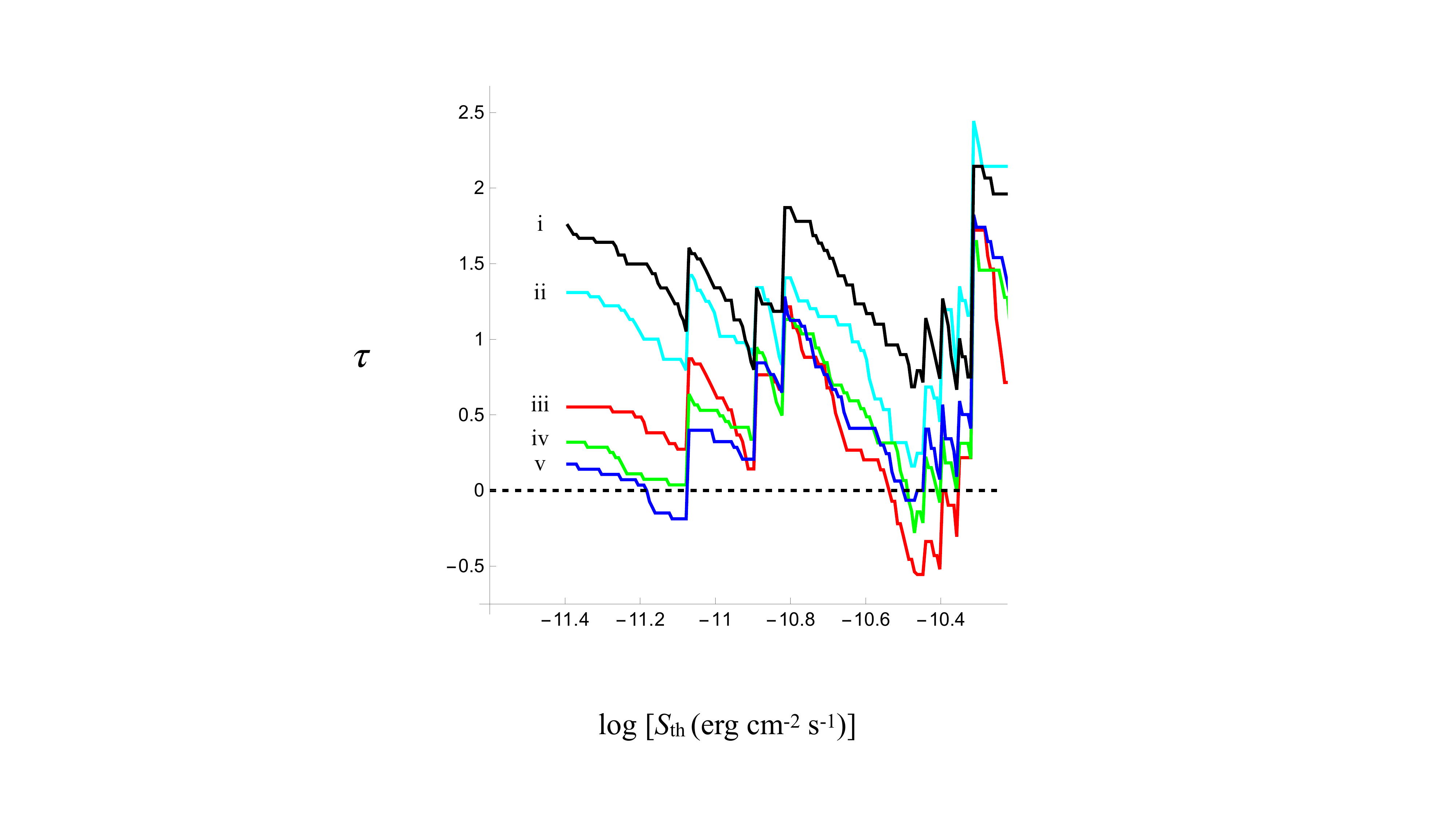}}
\caption{(a) The Efron--Petrosian statistic $\tau$ versus the logarithm of the flux threshold $S_{\rm th}$, over the interval $-11.4\le\log S_{\rm th}\le-10.25$, for the following values of $\alpha$: 2 (curve i coloured black), $1.75$ (curve ii coloured cyan), $1.5$ (curve iii coloured red), $1.25$ (curve iv coloured green) and 1.13 (curve v coloured blue).  Note that, in contrast to the curves for $\alpha=1.13$, $1.25$ and $1.5$ which respectively cross or closely approach the line $\tau=0$ at the flux thresholds $\log S_{\rm th}=-11.18$, $-10.49$ and $-10.54$ for the first time, the curves for $\alpha=1.75$ and $2$ lie well above the line $\tau=0$ throughout the interval $-11.4\le\log S_{\rm th}\le-10.25$.}
\label{UM3}
\end{figure}

\begin{figure}
\centerline{\includegraphics[width=12cm]{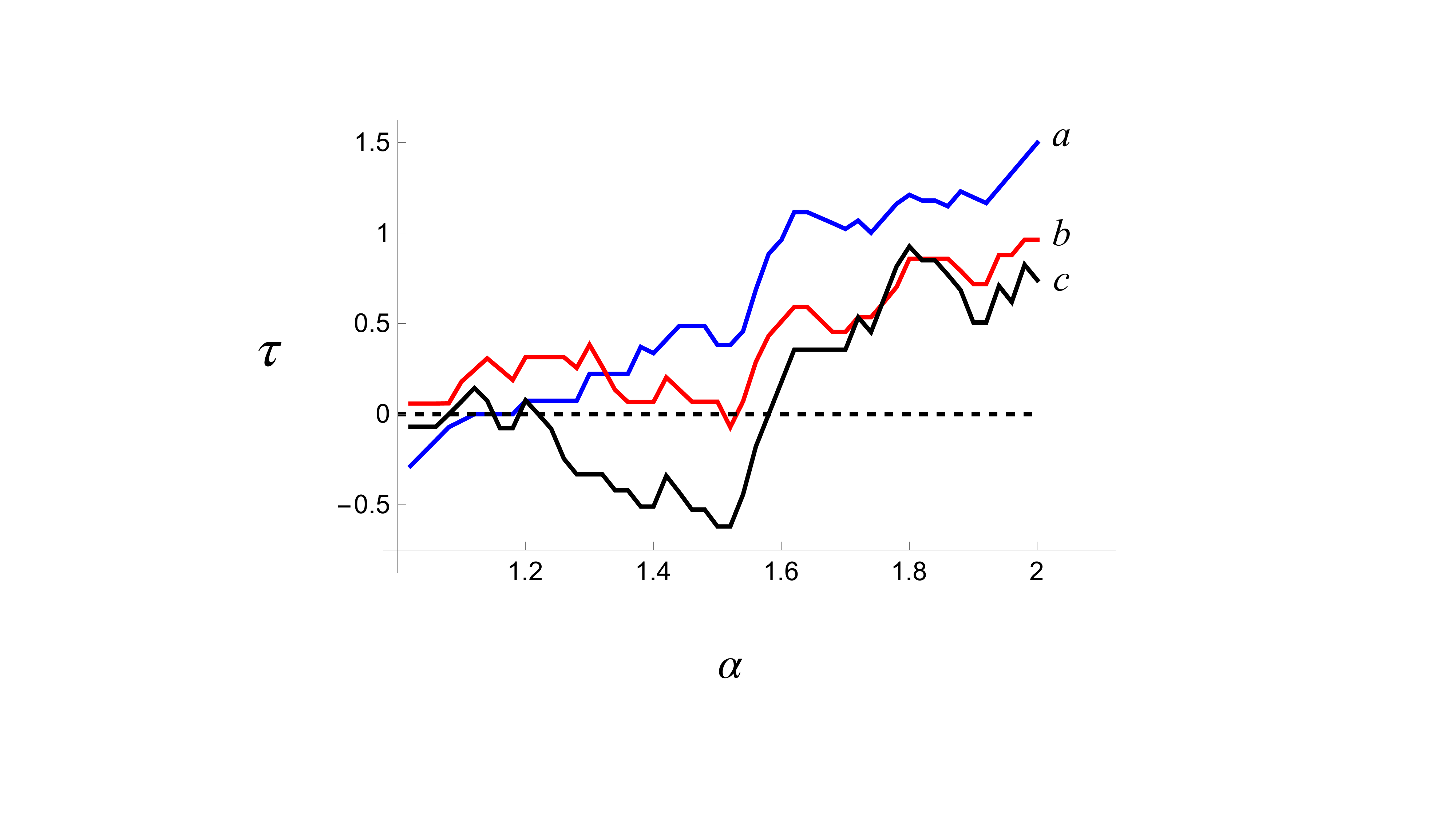}}
\caption{The Efron-Petrosian statistic $\tau$ as a function of the exponent $\alpha$ (in the dependence $S\propto D^{-\alpha}$ of flux density on distance) for the flux thresholds $a$, $b$ and $c$ designated by the dashed lines (with the same colours) in Fig.~\ref{UM1}.} 
\label{UM4}
\end{figure}

\subsection{Testing the hypothesis of independence of luminosities and distances of magnetars by the Efron-Petrosian method}
\label{subsec:results}

The expression in equation~(5) of~\citet{Ardavan2023a} for the Efron--Petrosian statistic $\tau$ is plotted as a function of the flux threshold $S_{\rm th}$ in Fig.~\ref{UM3} for five values of the exponent $\alpha$ in the dependence $S\propto D^{-\alpha}$ of flux density on distance: for $\alpha=2$ (curve i coloured black), $1.75$ (curve ii coloured cyan), $1.5$ (curve iii coloured red), $1.25$ (curve iv coloured green) and 1.13 (curve v coloured blue).  The corresponding dependence of the Efron-Petrosian statistic $\tau$ on the exponent $\alpha$ at fixed values of the flux threshold $S_{\rm th}$ is shown in Fig.~\ref{UM4} for $\log S_{\rm th}=-11.18$ (curve $a$ coloured blue), -10.54 (curve $b$ coloured red) and -10.4 (curve $c$ coloured black), i.e. for the flux thresholds designated by the dashed lines in Fig.~\ref{UM1}.

Note that the curve for $\alpha=2$ in Fig.~\ref{UM3} lies entirely above $\tau=0.67$: the value attained by the Efron-Petrosian statistic at the flux threshold $\log S_{\rm th}=-10.36$ which excludes $7$ of the $20$ elements of the data set depicted in Fig.~\ref{UM1}.  According to equation~(\ref{E7}), therefore, the hypothesis of independence of luminosity and distance for $\alpha=2$ can be rejected at any significance level $p$ exceeding $0.50$ even with the choice of a flux threshold ($\log S_{\rm th}=-10.36$) for which this significance level has its maximum value.  In other words, the significance levels above which this hypothesis can be rejected in the case of $\alpha=2$ are smaller than $p=0.50$ for any other flux threshold in the interval $-11.4\le\log S_{\rm th}\le-10.25$.  Likewise, the fact that the curve for $\alpha=1.75$ in Fig.~\ref{UM3} lies entirely above $\tau=0.16$ (the value it attains at the flux threshold $\log S_{\rm th}=-10.47$ which excludes $4$ of the $20$ elements of the present data set) implies that the hypothesis of independence of luminosity and distance can be rejected in the case of $\alpha=1.75$ at any significance level exceeding $0.87$ even with the choice of a flux threshold for which this significance level has its maximum value.

In contrast, there is a threshold excluding only one element of the data set ($\log S_{\rm th}=-11.18$) at which the curve for $\alpha=1.13$ (the blue curve v) intersects the line $\tau=0$ (the black dashed line) in Fig.~\ref{UM3}.  As already expected from the value $-1.13\pm0.05$ of the slope of the line fitted to the uncut data set in Fig.~\ref{UM1}b, the vanishing of the Efron-Petrosian statistic for $\alpha=1.13$ at the flux threshold $\log S_{\rm th}=-11.18$ (where the dashed lines $a$ are placed in Figs~\ref{UM1}a and~\ref{UM1}b) implies that the hypothesis of independence of luminosity and distance for $\alpha=1.13$ could not be rejected even at a $100\%$ confidence level (i.e.\ $p=1$) if we assumed that the exclusion of the single data point ($3.95, -11.59)$ below the dashed line $a$ in Fig.~\ref{UM1}b would result in a complete data set (see equation~\ref{E7}).  

This is an artefact of underestimating the flux threshold and confirms that the value of flux below which the data are incomplete lies closer to the peak of the data's histogram than the detection threshold $a$ of Fig.~\ref{UM1}a does~\citep[see][]{Bryant}.  In other words, the reason the low value ($1.13$) of the decay exponent $\alpha$ is not rejected by the Efron-Petrosian method when the truncation limit is set at the dashed line $a$ in Fig.~\ref{UM1} is that choosing this flux threshold is tantamount to making the erroneous assumption that the essentially uncut data set above dashed line $a$ is complete.

The same is true for the value $1.25$ of the decay exponent $\alpha$.  As indicated by the green curve iv in Fig.~\ref{UM3}, the Efron-Petrosian statistic assumes the value $\tau=0.037$  over the interval $-11.2\le \log S_{\rm th} \le -11.07$ in this case.  For the flux thresholds in this interval, the hypothesis of independence of luminosity and distance can be rejected only at significance levels exceeding $p=0.97$ (see equation~\ref{E7}).  This result, too, is an artefact of our having chosen too low a value of $S_{\rm th}$: the flux thresholds in the interval $-11.2\le \log S_{\rm th} \le -11.07$ only exclude a single element of the uncut data set in Fig.~\ref{UM1}.  There is no reason for assuming that the $19$-element data set thus obtained is more complete than the uncut $20$-element data set. 

\begin{figure}
\centerline{\includegraphics[width=10cm]{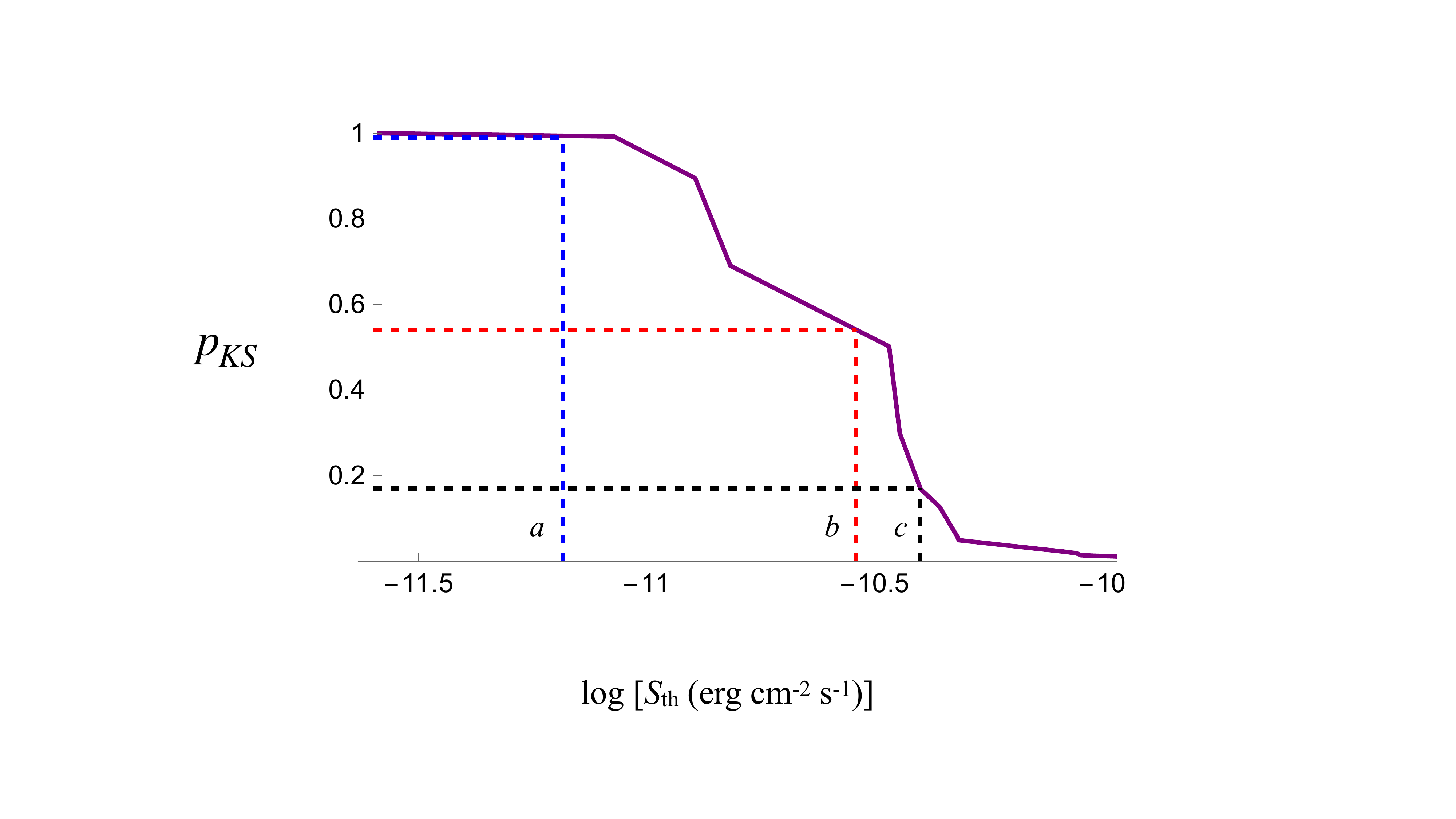}}
\caption{The Kolmogorov--Smirnov statistic $p_{KS}$ for testing whether the uncut $20$-element data set shown in Fig.~\ref{UM1} and a truncated version of it that only contains the elements with fluxes $S\ge S_{\rm th}$ are drawn from the same (unknown) distribution.  The values of $p_{KS}$ for the thresholds $\log S_{\rm th}= -11.8$, $-10.54$ and $-10.4$ (which here and in Fig.~\ref{UM1} are marked by the dashed lines $a$, $b$ and $c$) are $0.99$, $0.54$ and $0.17$, respectively.} 
\label{UM5}
\end{figure}

There is an upper limit, on the other hand, to how high the chosen value of the flux threshold $S_{\rm th}$ can be.  It is essential that the observationally obtained data set and the part of it that lies above the chosen flux threshold could be regarded as drawn from the same distribution: from the unknown distribution that is complete over all values of the flux density.  The Kolmogorov-Smirnov statistic $p_{KS}$ yields the probability that the truncated data set with the elements $S\ge S_{\rm th}$ and the uncut $20$-element data set depicted in Fig.~\ref{UM1} are drawn from the same (unknown) distribution.  For the present data set, the Kolmogorov-Smirnov statistic has the dependence shown in Fig.~\ref{UM5} on the flux threshold.

The vanishing of the Efron-Petrosian statistic for $\alpha=1.5$ at the flux threshold $\log S_{\rm th}=-10.54$ (where curve iii first crosses the line $\tau=0$ in Fig.~\ref{UM3} and where the dashed lines $b$ appear in Figs~\ref{UM1}a and~\ref{UM1}b) implies that the hypothesis of independence of luminosity and distance for $\alpha=1.5$ cannot be rejected even at a $100\%$ confidence level if the data set consisting of the data points above the dashed line $b$ in Fig.~\ref{UM1}b is regarded as complete (see equation~\ref{E7}).  In addition, the Kolmogorov--Smirnov test shows that the probability that the uncut $20$-element data set and the $16$-element truncated data set whose elements lie above the dashed line $b$ in Fig.~\ref{UM1}b are drawn from the same (unknown) distribution has the value $p_{KS}=0.54$ (see Fig.~\ref{UM5}).  

The above results can also be inferred from the plots, shown in Fig.~\ref{UM4}, of $\tau$ versus $\alpha$ at fixed values of $S_{\rm th}$.  The blue curve $a$ in Fig.~\ref{UM4} which corresponds to the flux threshold marked by the blue dashed line $a$ in Fig.~\ref{UM1}a coincides with the line $\tau=0$ over the interval $1.12\le\alpha\le1.18$ thereby implying that, given the choice $\log S_{\rm th}=-11.18$ of flux threshold, the hypothesis of independence of luminosity and distance cannot be rejected at any significance level if $\alpha$ lies in the interval $(1.12,1.18)$.  Likewise, the red curve $b$ in Fig.~\ref{UM4} which corresponds to the flux threshold marked by the red dashed line $b$ in Fig.~\ref{UM1}a first crosses the line $\tau=0$ at $\alpha=1.50$ thereby implying that, given the choice $\log S_{\rm th}=-10.54$ of flux threshold, the hypothesis of independence of luminosity and distance cannot be rejected even at a \%100 significance level if $\alpha$ equals $1.50$.

It would be possible to obtain a value of $\alpha$ larger than $1.50$ (but smaller than $1.75$) by choosing a higher flux threshold: the black curve $c$ in Fig.~\ref{UM4} which corresponds to the flux threshold marked by the black dashed line $c$ in Fig.~\ref{UM1}a crosses the line $\tau=0$ at $\alpha=1.58$ thereby implying that, given the choice $\log S_{\rm th}=-10.4$ of flux threshold, the hypothesis of independence of luminosity and distance cannot be rejected at any significance level if $\alpha=1.58$.  However, the probability that the resulting truncated data set has the same origin as the original uncut data set is significantly reduced at this higher value of the flux threshold.  The Kolmogorov-Smirnov statistic for the two data sets whose elements lie above the flux thresholds $b$ and $c$ has the values $p_{KS}=0.54$ and $0.17$, respectively (see Fig.~\ref{UM5}).  It is significantly less likely, therefore, that the observationally obtained data set and the part of it that lies above the flux threshold $c$ are drawn from the same distribution: from the unknown distribution that is complete over all values of the flux density.

Figure~\ref{UM5} shows, moreover, that $\tau$ at $\alpha=2$ has the values $0.96$ and $0.74$ for the choices $b$ ($\log S_{\rm th}= -11.8$) and $c$ ($\log S_{\rm th}= -10.4$) of the flux threshold, respectively.  The hypothesis of independence of luminosity and distance in the case of $\alpha=2$ can accordingly be rejected at any significance levels exceeding $\% 34$ and $\% 46$ for the flux thresholds $b$ and $c$, respectively (see equation~\ref{E7}).  In contrast, the corresponding hypothesis for the values $1.50$ and $1.58$ of $\alpha$ cannot be rejected even at a $\%100$ significance level with the same choices ($b$ and $c$ respectively) of the flux threshold.

The above results are in agreement with those found earlier~\citep[in][]{Ardavan2022a} by analysing the version of the data on magnetars' X-ray fluxes that is reported in the McGill Online Magnetar Catalog.\footnote{\url{http://www.physics.mcgill.ca/~pulsar/magnetar/main.html}}  Estimates of the effects both of observational errors and of the limited size of the data set in the Magnetar Outburst Online Catalogue on the present test results by the methods outlined in sections 3.4 and 3.5 of~\citet{Ardavan2023a} confirm, moreover, that neither of these effects are large enough to alter the conclusions reached in this section.

\subsection{Comparing magnetars' X-ray luminosities with their spin-down luminosities}
\label{sec:Xrayluminosity}

It follows from the results in Section~\ref{subsec:results} that the observational data in~\citet{MOOC2018} are consistent with the dependence $S\propto D^{-3/2}$ of the flux densities $S$ of magnetars on their distances $D$ at substantially higher levels of significance than they are with the dependence $S\propto D^{-2}$.  This does not contravene the requirements of the conservation of energy because the radiation that is generated by the magnetospheric current sheet is intrinsically transient.  The flux density of a steady-state emission decays as the inverse square of distance since the power that propagates across any two concentric spheres centred at the source is the same at all times.  In the present case, on the other hand, the rate of change of the energy density of the radiation with time is negative (instead of being zero as in a steady-state emission) so that the difference between the flux of power across two spheres centred at the source is compensated by the change with time of the energy contained inside the shell bounded by them~\citep[see][ appendix~C]{Ardavan_JPP}.

The factor by which the X-ray luminosity of a magnetar that is located at the distance $D$ is over-estimated when one uses the inverse-square law instead of $S\propto D^{-3/2}$ is $(D/\ell)^{1/2}$, in which the length $\ell$ is comparable to the light-cylinder radius of a neutron star~\citep[see][section~4]{Ardavan2023a}.  For a magnetar whose distance and rotation period have the values $8$ kpc and $5$ s, respectively, the order of magnitude of this factor is $10^6$.  If (as suggested by the fraction of known neutron stars that are identified as magnetars) we assume that the latitudinal beam-width of the X-ray emission from a magnetar is by the factor $10^{-2}$ smaller than that of the radio emission from most pulsars, the order of magnitude of the over-estimation factor reduces to $10^4$.   According to tables~1 and 3 of~\citet{MOOC2018}, however, the ratio of X-ray to spin-down luminosities (as inferred from the inverse-square law) is smaller than $10^4$ for every one of the listed magnetars.

The prevailing view that magnetars' X-ray luminosities exceed their spin-down luminosities~\citep{Mereghetti2015,Turolla2015,Kaspi2017,Borghese2019,Esposito2021} is not therefore upheld by the observational data in~\citet{MOOC2018}.  The X-ray luminosities of magnetars are over-estimated by the inverse-square law because magnetars, like gamma-ray pulsars, are observed along directions in which the radiation from the current sheet in their magnetospheres decays non-spherically.  The commonly accepted notion that radio-loud gamma-ray pulsars have gamma-ray luminosities that exceed their radio luminosities by several orders of magnitude~\citep{Lyne2022} has similarly been called into question by the results of an analysis of the data in the Fermi-LAT 12-Year Catalog~\citep[see][]{Ardavan2023a}.  Once the over-estimation of their values is rectified, the luminosities of gamma-ray pulsars, too, turn out to have the same range of values as do the luminosities of radio pulsars.  

Whereas the angle between the latitudinal direction in which the radiation is focused and the spin axis of a gamma-ray pulsar is normally fixed in time, this angle changes abruptly in the case of a flaring magnetar.  In other words, instead of propagating past the observer periodically, as in the case of a gamma-ray pulsar, the non-spherically decaying beam of radiation that is generated by the magnetospheric current sheet of a magnetar coincides with the observer's line of sight sporadically.  As a result, while the high-frequency radiation we receive from a gamma-ray pulsar is normally in the form of regular periodic pulses, that which we receive from a flaring magnetar is in the form of sporadic outbursts.  The observed differences between the spectra of magnetars and those of gamma-ray pulsars, moreover, have to do with the degree of proximity of the line of sight to one of the privileged directions into which their high-frequency radiation is beamed.  A magnetar's X-ray outbursts are expected to be accompanied by gamma-ray bursts in cases where the observer's line of sight lies sufficiently close to one of these privileged directions.

\section{X-ray and Gamma-ray spectra of magnetars}
\label{sec:spectra}

\subsection{SED of the caustics generated by the current sheet}
\label{subsec:spectrum}

The expression in equation (177) of~\citet{Ardavan2021} for the spectral distribution of the Poynting flux of the radiation from the magnetospheric current sheet can be written as
\begin{equation}
S_\nu=\kappa_0\, k^{-2/3}\left\vert\boldsymbol{\cal P}_2\, {\rm Ai}(-k^{2/3}\sigma_{21}^2)-{\rm i}k^{-1/3}\boldsymbol{\cal Q}_2\,{\rm Ai}^\prime(-k^{2/3}\sigma_{21}^2)\right\vert^2,
\label{E9}
\end{equation}
when the quantities ${\bar{\boldsymbol{\cal P}}}_l$, ${\bar{\boldsymbol{\cal Q}}}_l$ in that equation are negligibly smaller than their counterparts $\boldsymbol{\cal P}_l$ and $\boldsymbol{\cal Q}_l$ and the subscript $l$ has the value~$2$.  Here, Ai and ${\rm Ai}^\prime$ are the Airy function and the derivative of the Airy function with respect to its argument, the integer $k$ denotes the harmonic number $k=2\pi\nu/\omega$ associated with the radiation frequency $\nu$ and the angular frequency of rotation of the neutron star $\omega$, and $\kappa_0$ and $\sigma_{21}$ are two positive constants.  The coefficients of the Airy functions stand for $\boldsymbol{\cal P}_2=k^{-1/2}\boldsymbol{\cal P}_2^{(2)}$ and $\boldsymbol{\cal Q}_2=k^{-1/2}\boldsymbol{\cal Q}_2^{(2)}$ when $k \ge k_2$ and for $\boldsymbol{\cal P}_2=\boldsymbol{\cal P}_2^{(0)}$ and $\boldsymbol{\cal Q}_2=\boldsymbol{\cal Q}_2^{(0)}$ when $k < k_2$, in which the complex vectors $\boldsymbol{\cal P}_2^{(0)}$, $\boldsymbol{\cal Q}_2^{(0)}$, $\boldsymbol{\cal P}_2^{(2)}$ and $\boldsymbol{\cal Q}_2^{(2)}$ are defined by equations~(138)-(146) of~\citet{Ardavan2021} and $k_2$ designates a threshold frequency.  It is assumed here that both of the threshold frequencies, $k_1$ and $k_2$, that appear in equation~(138) of~\citet{Ardavan2021} are large and that $k_2 < k_1$.  The spectral distribution in equation~(\ref{E9}) is a characteristic feature of any radiation that entails caustics~\citep[see][]{Stamnes1986}.

Once equation~(\ref{E9}) is multiplied by the radiation frequency $\nu=\omega k/2\pi$ and $\boldsymbol{\cal P}_2$ and $\boldsymbol{\cal Q}_2$ are expressed in terms of $\boldsymbol{\cal P}_2^{(0)}$, $\boldsymbol{\cal Q}_2^{(0)}$, $\boldsymbol{\cal P}_2^{(2)}$ and $\boldsymbol{\cal Q}_2^{(2)}$, it yields
\begin{eqnarray}
\nu\, S_\nu&=&\frac{\omega\kappa_0}{2\pi}k^{1/3-j/2}\left\vert\boldsymbol{\cal P}^{(j)}_2\, {\rm Ai}(-k^{2/3}\sigma_{21}^2)\right.\nonumber\\*
&&\left.-{\rm i}k^{-1/3}\boldsymbol{\cal Q}^{(j)}_2\,{\rm Ai}^\prime(-k^{2/3}\sigma_{21}^2)\right\vert^2,
\label{E10}
\end{eqnarray}
where $j=0$ when $k<k_2$ and $j=2$ when $k\ge k_2$.  Squaring the complex vector inside the absolute-value signs, we arrive at
\begin{eqnarray}
\nu\, S_\nu&=&\kappa_1\, k^{1/3-j/2}\Big[{\rm Ai}^2(-k^{2/3}\sigma_{21}^2)+\zeta_1^2k^{-2/3}{{\rm Ai}^\prime}^2(-k^{2/3}\sigma_{21}^2)\nonumber\\*
&&+2\zeta_1\cos\beta\, k^{-1/3}{\rm Ai}(-k^{2/3}\sigma_{21}^2){\rm Ai}^\prime(-k^{2/3}\sigma_{21}^2)\Big],
\label{E11}
\end{eqnarray}
where
\begin{equation}
\kappa_1=\frac{\omega\kappa_0}{2\pi}\left\vert\boldsymbol{\cal P}^{(j)}_2\right\vert^2,\,\, \zeta_1=\frac{\left\vert{\boldsymbol{\cal Q}^{(j)}_2}\right\vert}{\left\vert{\boldsymbol{\cal P}^{(j)}_2}\right\vert},\,\,\cos\beta=\frac{\Im\left(\boldsymbol{\cal Q}^{(j)}_2\cdot\boldsymbol{\cal P}^{(j)*}_2\right)}{\left\vert\boldsymbol{\cal Q}^{(j)}_2\right\vert \left\vert\boldsymbol{\cal P}^{(j)}_2\right\vert},
\label{E12}
\end{equation}
and $\Im{}$ and $*$ denote an imaginary part and the complex conjugate, respectively.

To obtain the SED of the radiation, we must now integrate $S_\nu$ with respect to $\sigma_{21}$ over a finite interval $\rho\sigma_0\le\sigma_{21}\le\sigma_0$ with $\sigma_0\ll1$ and $0\le\rho<1$ (see~\citealt[][section~2]{Ardavan2024b}).  The result (obtained by means of Mathematica) is
\begin{eqnarray}
{\cal F}_\nu&=&\nu\int_{\rho\sigma_0}^{\sigma_0}S_\nu\,{\rm d}\sigma_{21}\nonumber\\*
&=& \kappa\,\chi^{1-j/2}\left[f_1(\chi,\rho)+\frac{\zeta^2}{4\sqrt{3}}f_2(\chi,\rho)-\frac{\zeta\cos\beta}{2\sqrt{3}}f_3(\chi,\rho)\right],\nonumber\\*
\label{E13}
\end{eqnarray}
where
\begin{equation}
\kappa=\frac{2^{j/2-1}\sigma_0^{3j/2}}{3^{(j+1)/2}\pi^{3/2}}\kappa_1,\qquad \zeta=\sigma_0 \zeta_1,\qquad\chi=\frac{2\sigma_0^3k}{3}=\frac{4\pi\sigma_0^3\nu}{3\omega},
\label{E14}
\end{equation}
\begin{eqnarray}
f_1&=&\bigg[3\Gamma\left(\frac{7}{6}\right)\eta\chi^{-2/3}{}_2F_3
\left(\begin{matrix}
1/6&1/6&{}\\
1/3&2/3&7/6
\end{matrix}
;-\eta^6\chi^2
\right)\nonumber\\
&&+\pi^{1/2}\eta^3{}_2F_3
\left(\begin{matrix}
1/2&1/2&{}\\
2/3&4/3&3/2
\end{matrix}
\,;-\eta^6\chi^2
\right)\nonumber\\
&&+\frac{9}{20}\Gamma\left(\frac{5}{6}\right)\eta^5\chi^{2/3}{}_2F_3
\left(\begin{matrix}
5/6&5/6&{}\\
4/3&5/3&11/6
\end{matrix}
\,;-\eta^6\chi^2
\right)\bigg]_{\eta=\rho}^{\eta=1},\nonumber\\
\label{E15}
\end{eqnarray}
\begin{equation}
f_2=\eta\chi^{-4/3}\,{}_{24}G^{31}\left(-\eta^2\chi^{2/3},\frac{1}{3}\left\vert\,\begin{matrix}
5/6&7/6&{}&{}\\
0&2/3&4/3&-1/6
\end{matrix}\right)\right.\bigg\vert_{\eta=\rho}^{\eta=1},
\label{E16}
\end{equation}
\begin{equation}
f_3=\eta\chi^{-1}\,{}_{24}G^{31}\left(-\eta^2\chi^{2/3},\frac{1}{3}\left\vert\,\begin{matrix}
5/6&1/2&{}&{}\\
0&1/3&2/3&-1/6
\end{matrix}\right)\right.\bigg\vert_{\eta=\rho}^{\eta=1},
\label{E17}
\end{equation}
and ${}_2F_3$ and ${}_{24}G^{31}$ are respectively the generalised hypergeometric function~\citep[see][]{Olver} and the generalised Meijer G-Function.\footnote{{\url{https://mathworld.wolfram.com/Meijer G-Function.html}}}  

The above SED contains the following six parameters: $j$ which equals $0$ or $2$ depending on whether the dimensionless frequency $k$ lies below or above the threshold frequency $k_2$, the trio $\zeta$, $\beta$, $\rho$ whose values determine the shape of the spectral distribution, and the duo $\kappa$, $\sigma_0$ which determine the position of this distribution along the flux-density (${\cal F}_\nu$) and the frequency ($\nu$) axes.  We will connect the values of these parameters to the physical characteristics of the current sheet in Section~\ref{subsec:connection}.  

\begin{figure}
\centerline{\includegraphics[width=9.9cm]{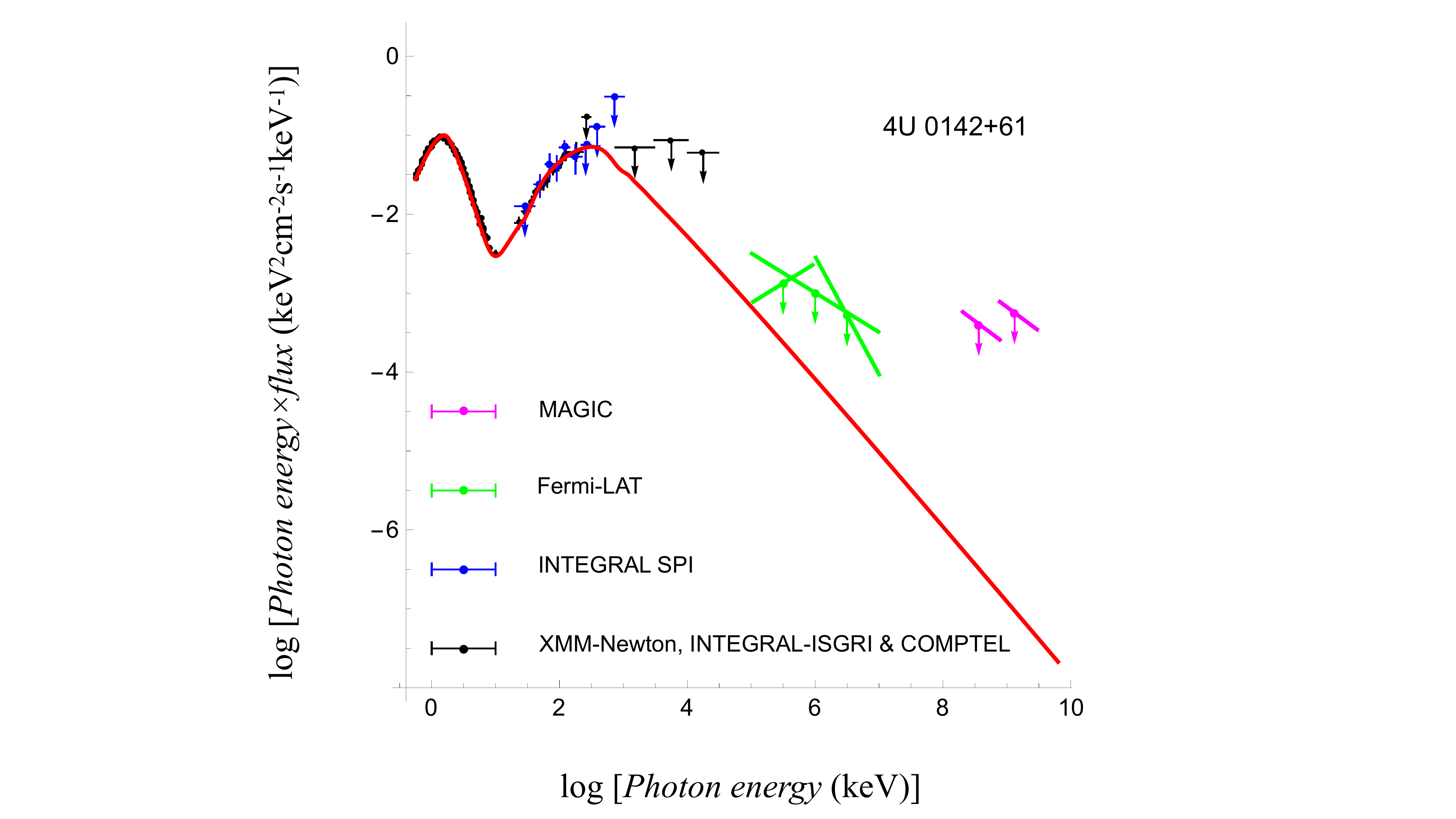}}
\caption{The unabsorbed total spectrum of the magnetar 4U 0142+61 from X-ray to TeV energies as measured with different instruments.  The curve (coloured red) is a plot of the SED described by equation~(\ref{E13}) for the parameters given in equations~(\ref{E18})--(\ref{E20}).}
\label{UM8}
\end{figure}

\subsection{Spectra of 4U0142+61, 1E1841-045 and XTE J1810-197 fitted with SED of the emission from their current sheet}
\label{subsec:fits}

In this section we determine the values of the six parameters appearing in the expression for ${\cal F}_\nu$ in equation~(\ref{E13}) for which this SED best fits the data on the high-energy spectra of each of the following three magnetars: 4U0142+61, 1E1841-045 and XTE J1810-197. 

\subsubsection{Spectrum of the magnetar 4U 0142+61}
\label{subsec:4U}

Figure~\ref{UM8} shows the unabsorbed total spectrum of 4U 0142+61 from X-ray to TeV energies as measured with different instruments: XMM-Newton ($0.55-11.5$ keV) and INTEGRAL-ISGRI ($20-300$ keV) in black, INTEGRAL SPI ($20-1000$ keV) in blue, CGRO COMPTEL ($0.75-30$ MeV) upper limits in black, Fermi-LAT ($0.1-10$ GeV) upper limits in green, and MAGIC ($> 200$ GeV) upper limits in magenta~\citep{denHartog2008,FermiMagnetars,MagicMagnetars}.  It also shows a plot of the SED described by equation~(\ref{E13}) that fits these data best.  

The parameters for which the function ${\cal F}_\nu(\chi)$ (depicted by the red curve in Fig.~\ref{UM8}) is plotted have the following values:
\begin{eqnarray}
\kappa&=&6.92\times10^{15}\quad{\rm keV}^2\,{\rm cm}^{-2}\,{\rm s}^{-1}\,{\rm keV}^{-1},\nonumber\\*
\chi&=&1.02\times(h\nu)_{\rm keV},\,\,\, j=0,\nonumber\\*
\sigma_0&=&8.99\times10^{-7},\,\,\, \zeta=1.05,\nonumber\\*
\beta&=&0,\quad\rho=1-10^{-18},
\label{E18}
\end{eqnarray}
over $0.55\le (h\nu)_{\rm keV}\le2.02$,
\begin{eqnarray}
\kappa&=&1.26\times10^{-2}\quad{\rm keV}^2\,{\rm cm}^{-2}\,{\rm s}^{-1}\,{\rm keV}^{-1},\nonumber\\*
\chi&=&1.32\times10^{-2}\times(h\nu)_{\rm keV},\,\,\, j=2,\nonumber\\*
\sigma_0&=&2.11\times10^{-7},\,\,\, \zeta=0.95,\nonumber\\*
\beta&=&0.275,\quad\rho=0.8,
\label{E19}
\end{eqnarray}
over $2.02\le (h\nu)_{\rm keV}\le28.2$, and
\begin{eqnarray}
\kappa&=&2.51\times10^{-2}\quad{\rm keV}^2\,{\rm cm}^{-2}\,{\rm s}^{-1}\,{\rm keV}^{-1},\nonumber\\*
\chi&=&4.27\times10^{-3}\times(h\nu)_{\rm keV},\,\,\, j=2,\nonumber\\*
\sigma_0&=&1.45\times10^{-7},\,\,\, \zeta=0.8,\nonumber\\*
\beta&=&0.13,\quad\rho=0,
\label{E20}
\end{eqnarray}
over $28.2\le (h\nu)_{\rm keV}\le10^{10}$, in which $2\pi/\omega$ has been set equal to the period of 4U 0142+61 (i.e.\ $8.69$ s), and $h$ and $(h\nu)_{\rm keV}$ stand for the Planck constant and photon energy in units of keV, respectively.  The photon energy $(h\nu)_{\rm keV}=2.02$, across which the values of the fit parameters change for the first time, corresponds to the threshold frequency $k_2$.  The photon energy $(h\nu)_{\rm keV}=28.2$, across which the values of the fit parameters change for a second time, corresponds to the threshold frequency $k_1$ in equation (138) of~\citet{Ardavan2021}: for $k\ge k_1$, the values of $\boldsymbol{\cal P}_2^{(2)}$ and $\boldsymbol{\cal Q}_2^{(2)}$ change to $\boldsymbol{\cal P}_2^{(1)}+\boldsymbol{\cal P}_2^{(2)}$ and $\boldsymbol{\cal Q}_2^{(1)}+\boldsymbol{\cal Q}_2^{(2)}$, respectively.  

The values of the free parameters in the above three ranges of photon energies are different because the degree with which the caustics generated by the present emission mechanism are focused depends on frequency.  This can be seen from equations~(\ref{E18})--(\ref{E20}) by noting that the value of $\sigma_0$ (and hence that of $\sigma_{21}$) decreases as frequency increases, i.e. the separation between the two nearby stationary points of the phases of the emitted waves decreases with increasing frequency.  Thus the difference between the expressions for ${\cal F}_\nu$ in different ranges of values of photon energy (and hence those of $\nu$ and $k$) stems from the difference in the ranges of validity of the approximations used in their derivations~\citep[see][section~4.3]{Ardavan2021}.

The distinctly small value of the parameter $\rho$ in equation~(\ref{E18}) reflects the limited range of values of the variable $\sigma_{21}$ that contributes to the Poynting flux in the frequency interval $0.55\le (h\nu)_{\rm keV}\le2.02$ (see equations~\ref{E9} and \ref{E13}).  It indicates that the part of the spectrum over this frequency interval arises from a focal region of the detected waves in which the separation between the two nearby stationary points of their phases, though finite as revealed by the non-zero value of $\sigma_0$, remains essentially constant (see Section~\ref{subsec:spectrum}).

\begin{figure}
\centerline{\includegraphics[width=10cm]{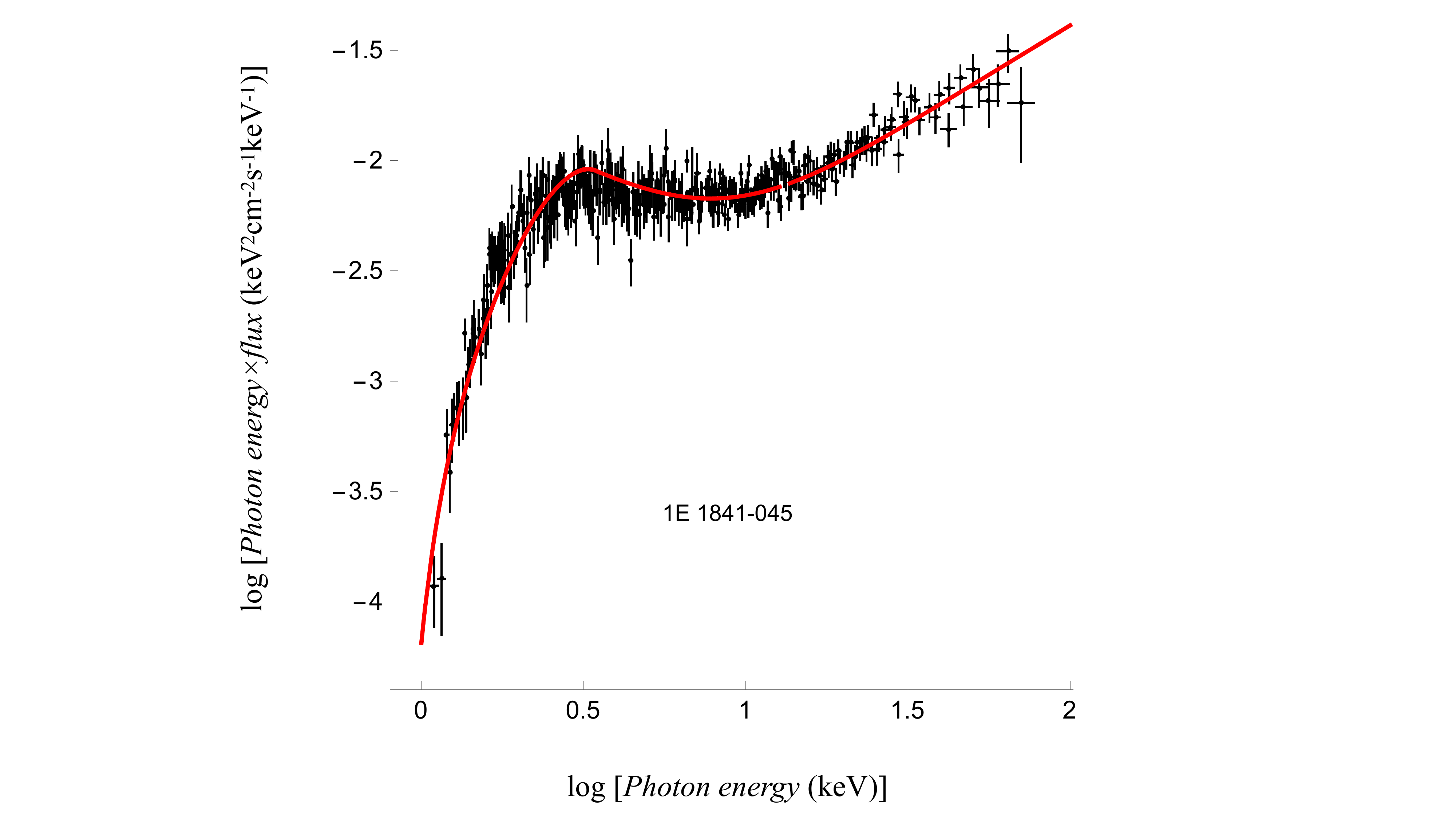}}
\caption{The phase-averaged spectrum of the magnetar 1E 1841-045 as measured by {\it Swift} XRT~\citep{An2013, Hascoet2014}.  The curve (coloured red) is a plot of the SED described by equation~(\ref{E13}) for the parameters given in equations~(\ref{E21}) and (\ref{E22}).}
\label{UM9}
\end{figure}

\subsubsection{Spectrum of the magnetar 1E 1841-045}
\label{subsec:1E} 

The data obtained by {\it Swift} XRT on the phase-averaged X-ray emission of the magnetar 1E 1841-045~\citep{An2013, Hascoet2014} are shown in Fig.~\ref{UM9}.  The SED that best fits these data (depicted by the red curve in Fig.~\ref{UM9}) is described by the expression in equation~(\ref{E13}) for the following values of its free parameters:
\begin{eqnarray}
\kappa&=&1.45\times10^{15}\quad{\rm keV}^2\,{\rm cm}^{-2}\,{\rm s}^{-1}{\rm keV}^{-1},\nonumber\\*
\chi&=&0.676\times(h\nu)_{\rm keV},\quad j=0,\nonumber\\*
\sigma_0&=&7.09\times10^{-7},\quad \zeta=10,\nonumber\\*
\beta&=&0,\quad\rho=1-10^{-20},
\label{E21}
\end{eqnarray}
over the range $1\le(h\nu)_{\rm keV}\le3.47$ and 
\begin{eqnarray}
\kappa&=&5.01\times10^{18}\quad{\rm keV}^2\,{\rm cm}^{-2}\,{\rm s}^{-1}{\rm keV}^{-1},\nonumber\\*
\chi&=&7.94\times10^{-4}\times(h\nu)_{\rm keV},\quad j=0,\nonumber\\*
\sigma_0&=&7.48\times10^{-8},\quad \zeta=0.3,\nonumber\\*
\beta&=&0.5,\quad\rho=1-10^{-20},
\label{E22}
\end{eqnarray}
over the range $3.47\le(h\nu)_{\rm keV}\le10^2$ of photon energies, in which $2\pi/\omega$ has been set equal to the period of 1E 1841-045 (i.e.\ $11.79$ s).  

As in the case of the spectrum of 4U 0142+61, the change in the values of the fit parameters across $(h\nu)_{\rm keV}=3.47$ arises from a change in the degree of focusing of the observed radiation: the difference between the two values of $\sigma_0$ ($7.20\times10^{-7}$ versus $8.07\times10^{-8}$) in the two different ranges of values of photon energy reflects a decrease in the separation between the two nearby stationary points of the phases of the emitted waves with increasing frequency~\citep[][sections~4.4 and 4.5]{Ardavan2021}. 

\subsubsection{Spectrum of the magnetar XTE J1810-197} 
\label{sebsec:XTE}

\begin{figure}
\centerline{\includegraphics[width=9cm]{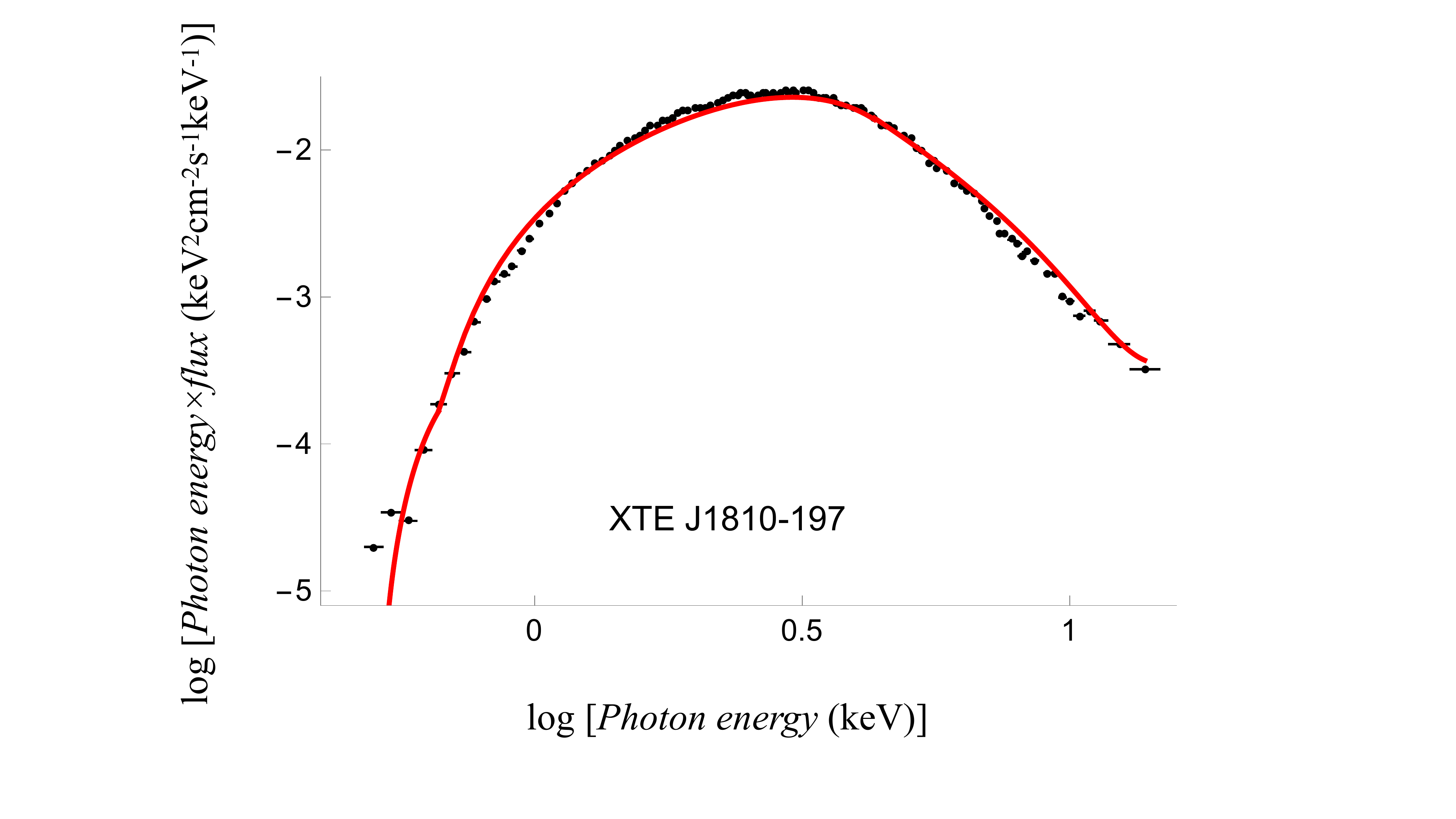}}
\caption{The unfolded spectrum of the magnetar XTE J1810-197 from the simultaneous {\it XMM-Newton} and {\it NuSTAR} observations performed in September 2019~\citep{BorgheseXTE}.  The curve (coloured red) is a plot of the SED described by equation~(\ref{E13}) for the parameters given in equations~(\ref{E23})--(\ref{E25}).}
\label{UM10}
\end{figure}

The simultaneous {\it XMM-Newton} and {\it NuSTAR} observations of the magnetar XTE J1810-197 performed in September 2019~\citep{BorgheseXTE} are shown in Fig.~\ref{UM10}.  The fit to these data (the red curve in Fig.~\ref{UM10}) is described by the expression in equation~(\ref{E13}) for the following values of its free parameters:
\begin{eqnarray}
\kappa&=&3.55\times10^{15}\quad{\rm keV}^2\,{\rm cm}^{-2}\,{\rm s}^{-1}\,{\rm keV}^{-1},\nonumber\\*
\chi&=&5.62\times(h\nu)_{\rm GeV},\quad j=0,\nonumber\\*
\sigma_0&=&1.85\times10^{-6},\quad \zeta=0.6\nonumber\\*
\beta&=&0,\quad\rho=1-10^{-20},
\label{E23}
\end{eqnarray}
over the range $0.525\le(h\nu)_{\rm keV}\le0.649$, 
\begin{eqnarray}
\kappa&=&8.91\times10^{-3}\quad{\rm keV}\,{\rm cm}^{-2}\,{\rm s}^{-1}\,{\rm keV}^{-1},\nonumber\\*
\chi&=&0.617\times(h\nu)_{\rm keV},\quad j=2,\nonumber\\*
\sigma_0&=&8.84\times10^{-7},\quad \zeta=2.4,\nonumber\\*
\beta&=&0,\quad\rho=0.87,
\label{E24}
\end{eqnarray}
over the range $0.661\le(h\nu)_{\rm keV}\le4.32$ and
\begin{eqnarray}
\kappa&=&5.31\times10^{-3}\quad{\rm keV}\,{\rm cm}^{-2}\,{\rm s}^{-1}\,{\rm keV}^{-1},\nonumber\\*
\chi&=&8.13\times10^{-3}\times(h\nu)_{\rm GeV},\quad j=2,\nonumber\\*
\sigma_0&=&2.09\times10^{-7},\quad \zeta=0.95,\nonumber\\*
\beta&=&0.14,\quad\rho=0.80,
\label{E25}
\end{eqnarray}
over the range $4.32\le(h\nu)_{\rm keV}\le13.8$ of photon energies, in which $2\pi/\omega$ has been set equal to the period of XTE J1810-197 (i.e.\ $5.54$ s). 

As in the case of the spectrum of 4U 0142+61, the photon energy $(h\nu)_{\rm keV}=0.661$, across which the values of the fit parameters change for the first time, corresponds to the threshold frequency $k_2$.  The photon energy $(h\nu)_{\rm keV}=4.32$, across which the values of the fit parameters change for a second time, corresponds to the threshold frequency $k_1$: for $k\ge k_1$, the values of $\boldsymbol{\cal P}_2^{(2)}$ and $\boldsymbol{\cal Q}_2^{(2)}$ change to $\boldsymbol{\cal P}_2^{(1)}+\boldsymbol{\cal P}_2^{(2)}$ and $\boldsymbol{\cal Q}_2^{(1)}+\boldsymbol{\cal Q}_2^{(2)}$, respectively~\citep[see][equation~138]{Ardavan2021}.  

\subsection{Inferring the characteristics of the source of the observed radiation from the values of the fit parameters of the SED}
\label{subsec:connection}

The free parameters of the present SED are related to the coefficients $\kappa_0$, $\boldsymbol{\cal P}_2^{(j)}$ and $\boldsymbol{\cal Q}_2^{(j)}$ that appear in the expression~(\ref{E10}) for the Poynting flux via equations~(\ref{E12}) and (\ref{E14}).  The coefficients $\kappa_0$, $\boldsymbol{\cal P}_2^{(j)}$ and $\boldsymbol{\cal Q}_2^{(j)}$ are in turn connected to the physical characteristics of the magnetospheric current sheet via the analysis in~\citet{Ardavan2021} by which the expression for the Poynting flux is derived.  In this section we express the values of these coefficients in terms of the inclination angle of the central neutron star, $\gamma$, and the following dimensionless versions of the magnitude $B_0$ of the star's magnetic field at its magnetic pole, distance $R_P$ of the observation point $P$ from the star, radius $r_{s0}$ of the star, and the rotation frequency $\omega$ of the star: 
\begin{equation}
{\hat B}_0=B_0/(10^{12} \,{\rm Gauss}),\quad D=R_P/(1\, {\rm kpc}),
\label{E26}
\end{equation}
\begin{equation}
d=r_{s0}/(10^6\,{\rm cm}), \quad {\hat P}=(10^2\,{\rm rad}\,{\rm s}^{-1})/\omega.
\label{E27}
\end{equation}
The spherical polar coordinates ($R_P, \varphi_P,\theta_P$) of the observation point $P$ are here defined with reference to a frame centred at the star whose $z$-axis lies along the rotation axis of the star.

From an analysis identical to that presented in~\citet[][section~4]{Ardavan2024b}, in which the parameter $\kappa$ is expressed in units of keV${}^2$~s${}^{-1}$~cm${}^{-2}$~keV${}^{-1}$ instead of erg~s${}^{-1}$~cm${}^{-2}$, it follows that 
\begin{eqnarray}
{\hat B}_0\,d^2&=&10^{-2}D\,{\hat P}\,\kappa^{1/2}\sigma_0^{3/2}(\Delta\chi)^{-1/2}\nonumber\\*
&&\times\left(1.68\,{\tilde\kappa}_{\rm th}^{-1/2}\delta_{j0}+2.57\,{\hat\kappa}_{\rm th}^{-1/2}\delta_{j2}\right),
\label{E28}
\end{eqnarray}
where $\delta_{j0}$ and $\delta_{j2}$ are Kronecker deltas and ${\tilde\kappa}_{\rm th}$ and ${\hat\kappa}_{\rm th}$ are two dimensionless functions whose dependences on the inclination angle $\gamma$ is plotted in Fig.~12 of~\citet{Ardavan2024a} for $\sigma_0\ll1$ and for an observation point in the far zone.

The dimensionless frequency $k=2\pi\nu/\omega$ of a $1$ keV photon is of the order of $10^{18}$ for a magnetar whose period is $5$ s.  So, the small values of the fit parameter $\sigma_0$ for which the present SED fits the data (see equations~\ref{E18}--\ref{E25}) imply that the corresponding value of $\chi=2\sigma_0^3k/3$ lies between $10^{-2}$ and $10^2$ for a $10$ keV X-ray photon.  Given that the data on fluxes are obtained by counting X-ray photons, it follows that the widths of the equivalent frequency bins over which flux is measured may be approximated by $\Delta k=3\sigma_0^{-3}\Delta\chi/2$ with a $\Delta\chi$ of the order of unity. 

It should be added that, as implied by the broadband nature of the polarization profiles depicted in Figs~8-16 of~\citet{Ardavan2021}, the observed polarization of magnetars' X-ray signals~\citep{Zane2023,Heyl2024,Taverna2024} is an intrinsic feature of the emission generated by the magnetospheric current sheet.

\subsubsection{Traits of the central neutron star and magnetosphere of the magnetar 4U 0142+61}
\label{subsec:4USource}

Given that the spectrum of the magnetar 4U~0142+61 over the photon energies $28.2$ to $10^{10}$ keV both covers a much wider range of photon energies than the other parts of the spectrum plotted in Fig.~\ref{UM8} and corresponds to the high-energy tail of the spectral distribution ${\cal F}_\nu$ where the asymptotic approximation used in deriving the expression given by equation~(\ref{E13}) describes this distribution most accurately, here we base our analysis of the properties of the central neutron star of this magnetar and its magnetosphere on the values of the fit parameters given in equation~(\ref{E20}).

Inserting the period, $8.69$ s, and the distance, $3.6$ kpc, of 4U~0142+61, together with the values of the fit parameters $j$ and $\kappa$ given in equation~(\ref{E20}), in equation~(\ref{E28}) we obtain
\begin{equation}
{\hat B}_0d^2=2.03\, {\hat\kappa}_{\rm th}^{-1/2}
\label{E44}
\end{equation}
for $\Delta\chi=1$.  If we assume that the star's magnetic field at its magnetic pole has the value $B_0=1.3\times10^{14}$ Gauss given by the conventional formula for magnetic dipole radiation, equation~(\ref{E44}) yields $\log{\hat\kappa}_{\rm th}^{-1/2}=1.81$ for a star of radius $10^6$ cm, i.e. for ${\hat B}_0=1.3\times10^2$ and $d=1$.  This value of ${\hat\kappa}_{\rm th}$ in conjunction with Figs~11 and 12 of~\citet{Ardavan2024a} would then imply that, in this case, the star's inclination angle $\gamma$ and the colatitude $\theta_{2PS}$ of the observation point have the values  $80.1^\circ$ and $147.5^\circ$, respectively.

Within the framework of the emission mechanism considered here, however, the value of ${\hat B}_0$ that is estimated by means of the formula for magnetic dipole radiation has no relevance to the value of the variable ${\hat B}_0$ that appears in equation~(\ref{E28}).  In the present context, the only constraint on the value of ${\hat B}_0 d^2$, when $j=2$, is that set by the inequality ${\hat\kappa}_{\rm th}\ge 1.42$ (see the curve delineated by the red dots in Fig.~12 of~\citealt{Ardavan2024a}): a constraint that, according to equation~(\ref{E44}), translates into ${\hat B}_0\ge2.88\,d^{-2}$ in the case of 4U~0142+61.  Hence, the order of magnitude of the magnetic field of the central neutron star of the magnetar 4U~0142+61 need not exceed $10^{12}$ Gauss when the radius of that star is of the order of $10^6$ cm.

\subsubsection{Traits of the central neutron star and magnetosphere of the magnetar 1E~1841-045}
\label{subsec:1ESource}

The period, $11.79$~s, and the distance, $8.5$~kpc, of 1E~1841-045 together with the value of the fit parameter $\kappa$ given in equation~(\ref{E22}), which applies to a wider range of frequencies than that given in equation~(\ref{E21}), yield
\begin{equation}
{\hat B}_0d^2=1.23\, {\tilde\kappa}_{\rm th}^{-1/2}
\label{E45}
\end{equation}
for $\Delta\chi=1$, since $j$ equals zero in this case (see equation~\ref{E28}).  According to the curve corresponding to $j=0$ in Fig.~12 of~\citet{Ardavan2024a}, the inclination angle $\gamma$ of this magnetar assumes the value $84.5^\circ$ if it is assumed that $r_{s0}=10^6$ cm and $B_0=6.9\times10^{14}$ Gauss (as implied by the conventional formula for magnetic dipole radiation).  Figure~11 of~\citet{Ardavan2024a} would in turn show that the colatitude $\theta_{P2S}$ along which 1E~1841-045 is observed has the value $161.8^\circ$.

One obtains radically different results, however, once one abandons the assumption that $B_0$ has the value implied by the conventional formula for magnetic dipole radiation.  In contrast to the curve pertaining to $j=2$ in Fig.~12 of~\citet{Ardavan2024a}, that for $j=0$ (which is delineated by the blue dots) does not set any constraint on the value of ${\hat B}_0 d^2$.  According to equation~(\ref{E45}) and Fig.~12 of~\citet{Ardavan2024a}, the magnitude $B_0$ of the magnetic field of the central neutron star of 1E~1841-045 at its magnetic pole would be less than $10^{12}$ Gauss for all inclination angles $0<\gamma\le 39.03^\circ$ if the radius of the neutron star equals $10^6$ cm.  

\subsubsection{Traits of the central neutron star and magnetosphere of the magnetar XTE J1810-197}
\label{subsec:XTESource}

For the same reasons as those pointed out in Section~\ref{subsec:4USource}, the characteristics of the central neutral star of XTE~J1810-197 are more accurately reflected in the fit parameters of its spectrum over the photon energies $4.32$ to $13.8$ keV.  The value of $\kappa$ given in equation~(\ref{E25}), which corresponds to this frequency interval, together with the period, $5.54$~s, and the distance, $3.5$~kpc, of XTE~J1810-197 yield
\begin{equation}
{\hat B}_0d^2=0.578\, {\hat\kappa}_{\rm th}^{-1/2}
\label{E46}
\end{equation}
for $\Delta\chi=1$.  If we assume that the star's magnetic field at its magnetic pole has the value $B_0=2.1\times10^{14}$ Gauss given by the conventional formula for magnetic dipole radiation, equation~(\ref{E46}) would yield $\log{\hat\kappa}_{\rm th}^{-1/2}=2.56$ for a star of radius $10^6$ cm, i.e. for ${\hat B}_0=2.1\times10^2$ and $d=1$.  This value of ${\hat\kappa}_{\rm th}$ in conjunction with Figs~11 and 12 of~~\citet{Ardavan2024a} would then imply that, in this case, the star's inclination angle $\gamma$ and the observer's colatitude $\theta_{2PS}$ have the values  $85.6^\circ$ and $165.4^\circ$, respectively.

On the other hand, the fact that the pulse profile of XTE~J1810-197 has only a single peak~\citep{BorgheseXTE} implies that the range of values of the inclination angle of this magnetar is limited to $\gamma\le60^\circ$~\cite[see][sections~4.4 and 5.1]{Ardavan2021}.  If we assume that the inclination angle of XTE~J1810-197 lies in the interval $20^\circ\le\gamma\le60^\circ$, then it follows from the curve delineated by the red dots in Fig.~12 of~\citet{Ardavan2024a} and equation~(\ref{E46}) that $0.82\le {\hat B}_0\le3.2$ when $d=1$.

\section{Interpretation of the flaring activity of magnetars}
\label{sec:conclusion}

We have shown why magnetars' X-ray luminosities do not exceed their spin-down luminosities (Section~\ref{sec:Xrayluminosity}) and have illustrated how the observed features of magnetars' X-ray spectra can be fitted with the spectral distribution function of the same non-thermal emission mechanism that is at play in rotation-powered pulsars (Section~\ref{subsec:fits}).  Our task in this section is to point out that the flaring activity that characterizes magnetars is another feature of the emission mechanism discussed in~\citet{Ardavan2021} that shows up only in the presence of large-scale timing anomalies.

There are two latitudinal directions in each hemisphere along which the radiation that is generated by the magnetospheric current sheet decays more slowly with distance than predicted by the inverse-square law.  At observation points away from these directions, the value of the decay exponent $\alpha$ that appears in $S\propto D^{-\alpha}$ changes from $3/2$ to $2$ over a latitudinal interval of the order of a radian.  But the beam-width of the caustics for which $\alpha$ equals $3/2$ occupies a much shorter latitudinal interval of the order of $(D\omega/c)^{-1}$, where $c/\omega$ is the radius of the star's light cylinder~\citep[][sections~4.4 and 5.5]{Ardavan2021}.  So, the solid angle centred on the neutron star within which the value of $\alpha$ significantly differs from $2$ is only a small fraction of $4\pi$.  Not only the propagation of the high-amplitude caustics but also that of the high-frequency radiation is thus confined to a small solid angle~\citep[see][]{Ardavan2023b}. 

Furthermore, within the framework of the present emission mechanism, mode changes arise from changes in the positions of the privileged directions along which $\alpha=3/2$~\citep[see][Figs 8-16]{Ardavan2021}.  The positions of these privileged direction, on the other hand, sensitively depend on the inclination angle $\gamma$ and on the rotation frequency $\omega$ of the central neutron star (which enters the analysis in~\citealt{Ardavan2021} through the scaling factor $c/\omega$).  It is hence predicted by the results arrived at in~\citet{Ardavan2021} that a large-scale timing anomaly changes the amplitude and spectrum, as well as the shapes of the observed pulses, simultaneously.

These findings together with the fact that magnetar outbursts are often accompanied by timing anomalies such as glitches, mode changes, or quakes~\citep{Archibald2020,Champion2020,Lower2023,Younes2023,Tsuzuki2024,Hu2024,Fisher2024} imply, therefore, that the flare activity in magnetars is caused by sudden changes in the orientation of their non-spherically decaying radiation beams relative to the line of sight.  As one of the privileged directions along which the radiation from the current sheet decays more slowly than predicted by the inverse-square law either swings past or oscillates across the line of sight, the amplitude of the observed radiation rises by the factor $(D/\ell)^{1/2}$ by which the values of the flux density in directions close to and far from a critical latitude differ from one another (see Section~\ref{sec:Xrayluminosity}).  In particular, the spectral evolution that accompanies an outburst stems from changes in the values of the parameters $\sigma_0$ and $\rho$ (in the expression for the SED described by equation~\ref{E13}) which sensitively depend on the colatitude of the observation point (see Section~\ref{subsec:spectrum}).

The observed differences between the traits of a magnetar and those of other pulsars (e.g.\ gamma-ray pulsars) can thus be attributed to variability versus constancy of the privileged directions along which the radiation from the current sheet decays non-spherically.  When the angles between the tightly focused caustics that are generated by the current sheet and the spin axis of the central neutron star are fixed in time, these radiation beams propagate past a set of favourably-positioned observers periodically, as in the case of a regularly pulsating gamma-ray pulsar.  But occasional abrupt changes of the latitudes along which these radiation beams propagate would be detected as X-ray outbursts or giant radio pulses by any observers within the paths of the moving beams, as in the cases of flaring magnetars or transient radio episodes.  

What can be deduced from the results of the analyses of the observational data in Sections~\ref{sec:test} and~\ref{sec:spectra} and from the interpretation of magnetars' flaring activity put forward in this section, therefore, is that the emission mechanism of magnetars is no different from that described in~\citet{Ardavan2021,Ardavan2022b} which applies to any class of non-aligned neutron stars, including rotation-powered pulsars.  Indeed, there are a number of magnetars that emit giant radio pulses, as well as exhibiting other pulsar-like features, and there are numerous pulsars that undergo magnetar-like outbursts~\citep{Gotthelf2019,Borghese2020,Israel2021,BorgheseXTE,Caleb2022,Rajwade2022,Chu2023,Bansal2023,Tsuzuki2024,Hu2024,Fisher2024}. 

Thr following final remarks are in order: 
\begin{description}
\item[(i)] 
Attempts at explaining the radiation from neutron stars has so far been focused mainly on mechanisms of acceleration of charged particles (see e.g. the references in~\citealt{Melrose2021} and~\citealt{HESS2023}): an approach spurred by the fact that, once the relevant version of this mechanism is identified, one can calculate the electric current density associated with the accelerating charged particles involved and thereby evaluate the classical expression for the retarded potential that describes the looked-for radiation.  In the analysis on which the present paper is based, however, we have evaluated the retarded potential, and hence the generated radiation field, using the macroscopic distribution of charge-current density that is already provided by the numerical computations of the structure of a non-aligned pulsar magnetosphere~\citep[][section~2]{Ardavan2021}.  Both the radiation field thus calculated and the electric and magnetic fields that pervade the pulsar magnetosphere are solutions of Maxwell's equations for the same charge-current distribution.  These two solutions are completely different, nevertheless, because they satisfy different boundary conditions: the initial-boundary conditions with which the structure of the pulsar magnetosphere is computed are radically different from those with which the retarded solution of Maxwell's equations (i.e. the solution describing the radiation from a prescribed distribution of charges and currents) is derived (see~\citealt[][section~6]{Ardavan2021} and~\citealt[][section~5]{Ardavan2024a}).  

Given that the superluminally moving distribution pattern of the current sheet in the magnetosphere of a non-aligned neutron star is created by the coordinated motion of aggregates of subluminally moving charged particles~\citep[see][]{GinzburgVL:vaveaa,BolotovskiiBM:VaveaD,BolotovskiiBM:Radbcm}, the motion of any of its constituent charged particles is too complicated to be taken into account individually.  Only the densities of charges and currents enter Maxwell's equations, on the other hand, so that the macroscopic charge-current distribution associated with this current sheet (which comprises different particles at different times) takes full account of the contributions toward the radiation that arise from the complicated motions of the charged particles comprising it.  Each volume element of the uniformly-rotating distribution pattern of the magnetospheric current sheet acts as a point-like source of emission whose field embraces a synergy between the superluminal version of the field of synchrotron radiation and the vacuum version of the field of \v{C}erenkov radiation~\citep[][section~3]{Ardavan2021}.  Once superposed to yield the emission from the entire volume of the source, the contributions from the volume elements of this distribution pattern that approach the observation point with the speed of light and zero acceleration at the retarded time interfere constructively and form caustics in certain latitudinal directions relative to the spin axis of the neutron star.  The waves that embody these caustics are more focused the farther they are from their source: as their distance from their source increases, two nearby stationary points of their phases draw closer to each other and eventually coalesce at infinity~\citep[][section~6]{Ardavan2021}.  
 \item[(ii)] 
Thickness of the current sheet, which sets a lower limit on the wavelength of the generated radiation, is dictated by microphysical processes that are not well understood: the standard Harris solution of the Vlasov-Maxwell equations~\citep{Harris} which is commonly used in analysing a current sheet is not applicable in the present case because the current sheet in the magnetosphere of a non-aligned neutron star moves faster than light and so has no rest frame.  Even in stationary or subluminally moving cases, there is no consensus on whether equilibrium current sheets in realistic geometries have finite or zero thickness~\citep{Klimchuk}.  The fact that the SED described by equation~(\ref{E13}) yields such good fits to the observed spectra of the magnetars analysed here corroborates the notion that, though necessarily volume-distributed~\citep{BolotovskiiBM:Radbcm}, the magnetospheric current sheet is sufficiently thin to generate X-rays and gamma rays~\citep[see][section~4.7]{Ardavan2021}. 
\item[(iii)]
The dependence of the SED described by equation~(\ref{E13}) on the magnitude $B_0$ of the star's magnetic field at its magnetic pole stems from the dependence on $B_0$ of the global distribution of charge-current density that is predicted by the numerical simulations of the magnetospheric structure of a non-aligned neutron star~\citep[see][section~2]{Ardavan2021}.  Estimates of $B_0$ in the cases of the magnetars 4U0142+61, 1E1841-045 and XTE J1810-197, inferred from the values of the fit parameters for their observed spectra, clearly demonstrate that the central neutron stars of magnetars need not be more strongly magnetized than those of normal pulsars (see Section~\ref{subsec:connection}).  The formula for the Poynting flux of an obliquely rotating magnetic dipole in vacuum, by means of which the strength of the magnetic field of a magnetar is normally estimated, is based on a rudimentary emission mechanism that has no relevance to the emission mechanism by which the superluminally moving current sheet in the magnetosphere of a non-aligned neutron star radiates~\citep{Ardavan2021,Ardavan2022b}.

\end{description} 

\section*{Data availability}

The data used in this paper are available in the public domain.

\bibliographystyle{mnras}
\bibliography{Magnetars2.bib}

\bsp	
\label{lastpage}
\end{document}